# Microwave photonic radar jamming and target detection integration based on advanced waveform editing, forwarding, and self-squaring reception

Fangyi Yang, Hang Yang, Xin Zhong, Taixia Shi, and Yang Chen*

Shanghai Key Laboratory of Multidimensional Information Processing, School of Communication and Electronic Engineering, East China Normal University, Shanghai 200241, China
*Correspondence to: ychen@ce.ecnu.edu.cn

**ABSTRACT**
The integrated radar and jamming (IRAJ) system provides a promising solution that meets the demands for miniaturization, integration, and multifunctionality in complex warfare environments. However, traditional electronic-domain IRAJ systems face limitations in operating frequency and bandwidth. In this paper, we propose and experimentally demonstrate a microwave photonic IRAJ system based on pseudo-random binary phase modulation and segmented frequency shifting. By modulating pseudo-random binary coding sequence and frequency-shifting signals onto linearly frequency-modulated (LFM) pulses, an IRAJ waveform is generated to achieve noise-like jamming against the adversary radar. To overcome the random π-phase jumps introduced by pseudo-random binary modulation in the de-chirped signal, a time-domain squaring operation is implemented during de-chirped reception, restoring the radar detection ability of our system and enabling accurate target sensing without prior knowledge of the coding sequence. Experimental results demonstrate that the system can generate IRAJ waveforms with a bandwidth of up to 4 GHz, covering both 10–28 GHz. The proposed system achieves effective jamming against adversary radars employing either de-chirped reception or pulse compression, with the generated jamming results exhibiting an irregular and random distribution of false targets. Meanwhile, the system maintains radar performance with a ranging error of around 5 cm and a radial velocity measurement error below 4 cm/s.

## 1. Introduction

With the rapid advancement of technology and the escalating complexity of modern warfare environments [1], [2], electronic warfare (EW) is evolving toward informatization and systematization. Among its key components, radar jamming techniques play a critical role in electronic countermeasures by transmitting interference signals to disrupt adversary's detection systems. Building upon traditional radar jamming solutions, digital radio frequency memory (DRFM) has emerged as a widely adopted technology [3]. By digitally sampling and storing the intercepted radar signals, DRFM enables time-delay [4], [5], [6], phase modulation [7], [8], and other signal-editing operations in the digital domain, providing the flexibility to generate various jamming waveforms. However, due to the limited sampling rates of analog-to-digital converters (ADCs) and digital-to-analog converters (DACs), DRFM suffers from a restricted operating frequency and a narrow instantaneous bandwidth—typically only 1 GHz. This makes it difficult to keep pace with the high-frequency, wide-bandwidth advancements of modern radars, thereby significantly limiting its applicable scenarios.

Microwave photonics, with its wide frequency range, flexible tunability, low loss, and enhanced electromagnetic immunity [9], offers a promising approach for radar jamming techniques to break through the above constraints of the electronic bottleneck. A photonic-assisted jamming signal generation scheme that combines photonic mixing with DRFM largely extended the working frequency to the Ka band [10]. However, the issue of limited instantaneous bandwidth persists, with the deployed DRFM systems exhibiting a 500 MHz instantaneous bandwidth. In contrast, full-optical radar jamming signal generation schemes, which process the intercepted radar signal in the optical domain, allow for greater working frequency and broader operating bandwidth. For instance, optical recirculating loops can be used to store and achieve time delay on radar signals [11], [12], together with Doppler frequency shift for generating deceptive jamming [13], [14], [15]. Additionally, based on Serrodyne optical frequency translation, frequency shifting can also be achieved by sawtooth waveforms [16], [17], [18], while phase modulation can be realized by controlling the bias voltage of the electro-optic modulator to adjust its bias point [19], [20]. Furthermore, various advanced jamming types have been proposed and combined. Specifically, comb spectrum modulation jamming (CSMJ) and interrupted sampling repeater jamming (ISRJ) can be jointly implemented by interruptedly sampling the radar signal with a periodic rectangular signal and modulating its frequency with an electric comb signal [21], [22]. ISRJ and cross-cosinusoidal phase-modulated jamming can be combined through polarization multiplexing [23]. However, a common issue of these methods is that the generated false targets often exhibit a regular and equally spaced distribution. To address this problem, the combination of CSMJ and chopping and interleaving jamming (C&IJ) was proposed to generate irregularly distributed false targets [24]. Despite this, the false targets still retain an equally spaced distribution.

In the increasingly complex warfare environment, operational platforms not only need to implement radar jamming to gain an advantage in electronic countermeasures but also rely on radar for target detection and tracking. To address this demand, integrated radar and jamming (IRAJ) systems based on signal sharing have emerged, by using one single waveform to simultaneously perform target detection and radar jamming, effectively reducing device redundancy, enhancing resource utilization, and alleviating mutual interference between radio frequency (RF) devices. The concept of signal sharing was first introduced by multiplying linear frequency modulation (LFM) signals with random pulse cluster signals to generate IRAJ waveforms [25]. To address the limited randomness of the IRAJ waveform proposed in [25], subsequent studies [26], [27], [28] focused on improving the randomness and detection performance of the shared waveform. However, most electronic IRAJ systems remain in the theoretical and simulation stages, lacking experimental validation. Moreover, due to the electronic bottleneck, these systems suffer from fixed operating frequency band and limited operating bandwidth. Since microwave photonic radar jamming methods offer significant advantages in high frequency, broadband tunability, and diverse jamming formats, the integration of radar detection and jamming through microwave photonics presents a promising solution. Nevertheless, despite its potential, IRAJ systems based on microwave photonics have rarely been explored.

In this work, to the best of our knowledge, we propose and experimentally demonstrate the first microwave photonic IRAJ system based on pseudo-random binary phase modulation and segmented frequency shifting. The designed IRAJ waveform with a bandwidth up to 4 GHz and flexible tunability covering 10–28 GHz can achieve coherent noise-like jamming against adversary's radars employing both de-chirped reception and pulse compression. In contrast to current microwave photonic radar jamming methods, the jamming result generated by our system exhibits a random noise-like distribution, reducing the likelihood of identification and counteraction by adversary's radar. Meanwhile, to achieve the radar detection using the IRAJ waveform, a self-squaring operation is proposed and performed at the receiver, which successfully restores the target sensing ability and enables high-accuracy ranging and velocity measurement without prior knowledge of the coding sequence. Experimental results validate the radar function, demonstrating a ranging error within 5 cm and a radial velocity measurement error not exceeding 4 cm/s.

## 2. Principle and experimental setup

The schematic diagram and experimental setup of the proposed microwave photonic IRAJ system are shown in Fig. 1. A continuous-wave (CW) optical signal centered at fc and generated from a laser diode (LD, ID Photonics CoBriteDX1-1-HC1-FA) is expressed as $E_c(t)=E_c\exp(j2\pi f_c t)$, where $E_c$ represents the amplitude of the optical signal. This optical signal passes through an optical isolator (ISO) and is then split by a 50:50 optical coupler (OC1). One output of OC1 is used as the optical carrier and injected into a dual-parallel Mach–Zehnder modulator (DP-MZM1, Fujitsu FTM-7961EX), where it is modulated by two phase-orthogonal electrical LFM signals. Acting as the intercepted adversary's radar signal, the LFM signal with a pulse repetition period of $T$ and a bandwidth of $B$, is generated by a four-channel arbitrary waveform generator (AWG1, Keysight M8195A) and applied to DP-MZM1 via a 90° electrical hybrid coupler. The LFM signal is given by

$$S_{\mathrm{LFM}}(t)=V_1\cos\left[2\pi\left(f_0 t+\frac{1}{2}kt^2\right)\right]\mathrm{rect}\left(\frac{t-T_\mathrm{p}/2}{T_\mathrm{p}}\right), t\in[0,T], \tag{1}$$

where $V_1$, $f_0$, and $T_\mathrm{p}$ are the amplitude, initial frequency, and pulse duration of the LFM signal, respectively, and $k=B/T_\mathrm{p}$ is the chirp rate of the LFM signal. The term $\mathrm{rect}\left[\left(t-T_\mathrm{p}/2\right)/T_\mathrm{p}\right]$ represents a time-shifted rectangular window function centered at $T_\mathrm{p}/2$ with a width of $T_\mathrm{p}$. The time-frequency relationship of the generated LFM signal is illustrated in Fig. 1(a). The optical carrier is carrier-suppressed single-sideband (CS-SSB) modulated by the LFM signal, with only the −1st-order LFM optical sideband retained. Under the small-signal modulation condition, the output of DP-MZM1 can be expressed as

$$\begin{aligned} E_{\text{DP-MZM1}}(t)=-\frac{1}{4}E_c\beta_1 V_1\mathrm{rect}\left(\frac{t-T_\mathrm{p}/2}{T_\mathrm{p}}\right) \\ \times\exp\left\{\mathrm{j}\left[2\pi\left(f_c-f_0\right)t-\pi kt^2\right]\right\} \end{aligned}, \tag{2}$$

where $\beta_1=\pi/V_{\pi1}$, and $V_{\pi1}$ denotes the half-wave voltage of DP-MZM1. Subsequently, the −1st-order LFM optical sideband from DP-MZM1 is fed into OC2.

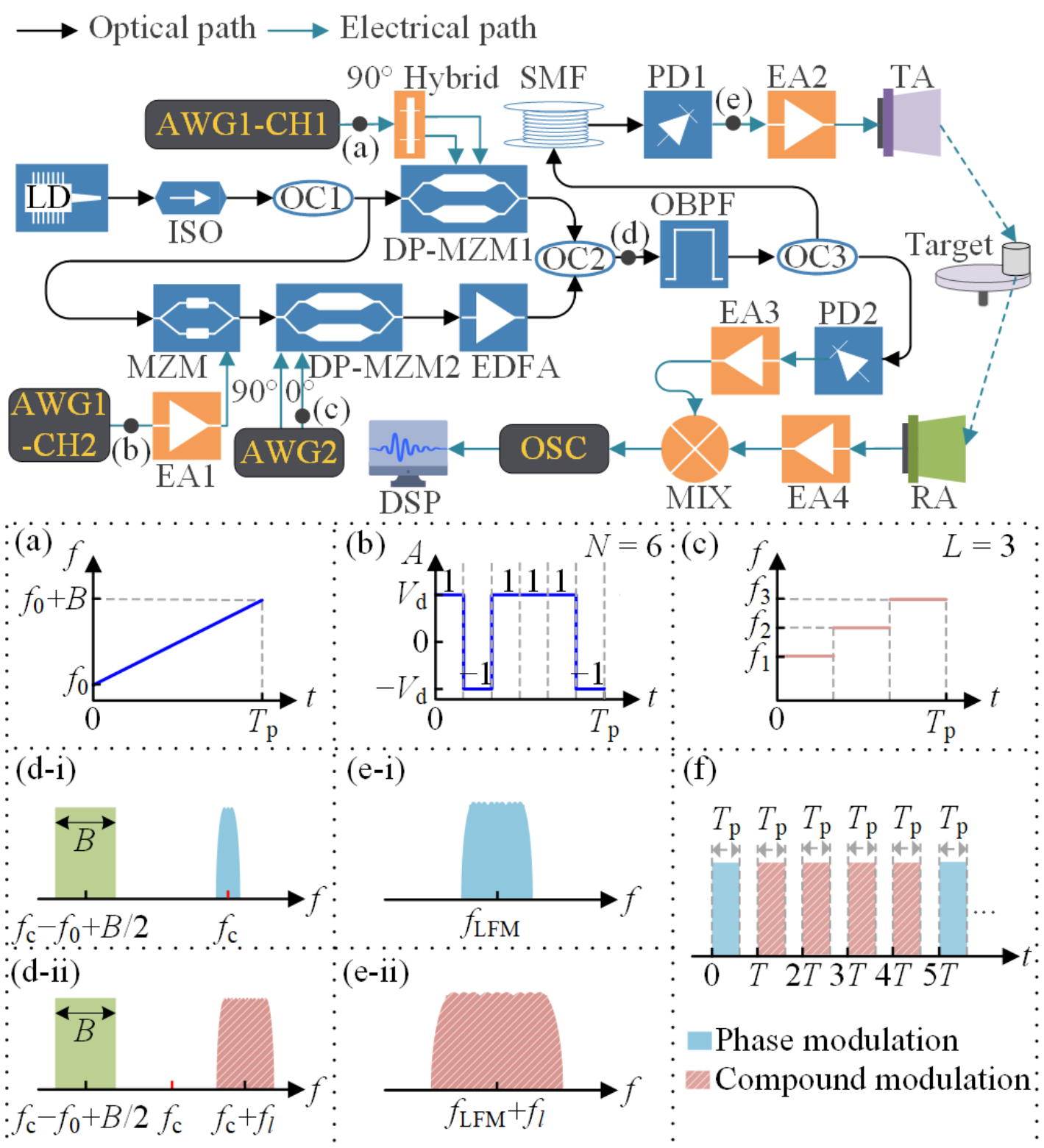

Fig. 1. Schematic diagram and experimental setup of the proposed integrated radar detection and jamming system. (a)–(e) are the diagrams of key nodes in the system diagram; (f) time-domain pulses of the generated IRAJ waveforms. LD, laser diode; ISO, isolator; OC, optical coupler; DP-MZM, dual-parallel Mach–Zehnder modulator; MZM, Mach–Zehnder modulator; SMF, single-mode fiber; EDFA, erbium-doped fiber amplifier; OBPF, optical band-pass filter; PD, photodetector; AWG, arbitrary waveform generator; EA, electrical amplifier; TA, transmitting antenna; RA, receiving antenna; MIX, mixer; OSC, oscilloscope; DSP, digital signal processing.

The other output of OC1 is sent to a Mach–Zehnder Modulator (MZM, Fujitsu FTM-7938EZ). A baseband binary signal $d(t)$ from AWG1 is amplified by an electrical amplifier (EA1, Multilink MTC5515) and applied to the MZM. The baseband binary signal $d(t)$ is based on a pre-generated pseudo-random binary sequence $c_n \in \{-1,1\}$, $n = 0,1,\ldots,N-1$, and $N$ represents the length of the binary sequence. Each code symbol has a duration of $T_\mathrm{d} = T_\mathrm{p}/N$, and the repetition period of the baseband binary signal is also $T$. Thus, $d(t)$ is denoted as

$$d(t) = V_\mathrm{d}\sum_{n=0}^{N-1} c_n \mathrm{rect}\left(\frac{t - nT_\mathrm{d} - T_\mathrm{d}/2}{T_\mathrm{d}}\right) \tag{3}$$

where $V_\mathrm{d}$ is the amplitude of the baseband binary signal $d(t)$, while the polarity of the binary phase-coded signal is determined by $c_n$. The waveform of $d(t)$ for $N = 6$ is illustrated in Fig. 1(b).

The output of the MZM can be expressed as

$$E_\mathrm{MZM}(t) = \frac{1}{2}E_\mathrm{c}(t)\cos\left(\frac{\pi V_\mathrm{dc}}{V_{\pi 2}} + \frac{\pi d(t)}{V_{\pi 2}}\right), \tag{4}$$

where $V_{dc}$ and $V_{\pi 2}$ are the bias voltage and the half-wave voltage of the MZM, respectively. When $V_{dc} = V_{\pi 2}/2$, (4) can be simplified to

$$E_{\mathrm{MZM}}(t) = -\frac{1}{2}\xi E_{\mathrm{c}}(t)\sum_{n=0}^{N-1} c_n \mathrm{rect}\left(\frac{t - nT_d - T_d/2}{T_d}\right), \tag{5}$$

where $\xi = \sin(\pi V_{\mathrm{d}}/V_{\pi 2})$ represents the amplitude factor of the output of MZM. The sign of the output optical carrier is dictated by $c_n$. Besides, to maximize the amplitude of the output of MZM, $V_{\mathrm{d}}$ is designed to be as close as possible to $V_{\pi 2}/2$. Accordingly, $\xi \approx 1$, and phase modulation of the optical carrier with two separate phases, i.e., 0 and π, can be realized.

Subsequently, the phase-modulated optical carrier is further processed by DP-MZM2 (Fujitsu FTM-7961EX), which is biased as a CS-SSB modulator, serving as an optical frequency shifter. Two phase-orthogonal RF signals, i.e., $S_{\text{I-RF}}(t)$ and $S_{\text{Q-RF}}(t)$, are input into the RF ports of DP-MZM2. To introduce complex frequency-shifting interference, the frequency of the RF signal can undergo multiple changes within a single period of the LFM signal. These RF signals are generated by AWG2 (RIGOL DG2052). The in-phase RF signal $S_{\text{I-RF}}(t)$ can be expressed as

$$E_{\mathrm{MZM}}(t) = -\frac{1}{2}\xi E_{\mathrm{c}}(t)\sum_{n=0}^{N-1} c_n \mathrm{rect}\left(\frac{t - nT_d - T_d/2}{T_d}\right), \tag{6}$$

$$\mathrm{rect}_l(t) = \mathrm{rect}\left[\frac{t - (l-1)T_{\mathrm{p}}/L - T_{\mathrm{p}}/(2L)}{T_{\mathrm{p}}/L}\right]. \tag{7}$$

Here, $\mathrm{rect}_l(t)$ denotes the rectangular window function for the $l$-th time segment. $V_{\mathrm{RF}}$ is the amplitude of the RF signal, $L$ is the number of frequency shift segments per pulse, and $\delta \in \{0,1\}$ is the frequency-shifting control factor. For $\delta = 1$, to achieve equal-time-interval frequency shifting of the LFM signal, the pulse width $T_{\mathrm{p}}$ is equally divided into $L$ segments, and different frequencies $f_l$ ($l$=1,2,…,$L$) are sequentially applied to the DP-MZM2. For $\delta = 0$, no frequency shifting is applied. Considering $\delta = 1$ and $L = 3$, the time-frequency diagram of the RF signal $S_{\text{I-RF}}(t)$ with equally spaced frequency increments is shown in Fig. 1(c). Note that each frequency $f_l$ can be flexibly set to realize customized frequency shifting of the optical carrier. Due to CS-SSB modulation, the output of DP-MZM2 is given by

$$E_{\mathrm{DPMZM2}}(t) = -\frac{1}{2}\beta_2 V_{\mathrm{RF}} E_{\mathrm{MZM}}(t)\sum_{l=1}^{L}\exp(\mathrm{j}2\pi\delta f_l t)\mathrm{rect}_l(t), \tag{8}$$

where $\beta_2 = \pi/V_{\pi 3}$, and $V_{\pi 3}$ is the half-wave voltage of DP-MZM2. It can be seen that after the cascaded modulation by the MZM and DP-MZM2, the optical carrier undergoes not only phase modulation but also segmented frequency shifting controlled by the factor $\delta$. Specifically, when $\delta$ is set as 0, only phase modulation exists, while when $\delta$ is set as 1, a compound modulation that combines phase modulation and frequency shifting is employed.

Subsequently, the output from DP-MZM2 is amplified by an erbium-doped fiber amplifier (EDFA, Amonics AEDFA-PA-35-B-FA) before being directed into OC2, making its optical power approximately equal to that of the LFM optical sideband in the upper branch. A dense wavelength division multiplexer is then used as an optical band-pass filter (OBPF) to filter out most of the amplified spontaneous emission noise. The spectra of the output optical signal at OC2, corresponding to the cases of $\delta = 0$ and $\delta = 1$, are shown in Fig. 1(d-i) and Fig. 1(d-ii), respectively. After being filtered by OBPF, the optical signal is then split into two parts by OC3. One part of the optical signal is injected into a 200 m single-mode fiber (SMF), and then the IRAJ waveform is generated by a photodetector (PD1, u2t XPDV2120RA), whose expression can be written as

$$i(t) \propto \gamma \sum_{n=0}^{N-1} c_n \mathrm{rect}_n(t) \sum_{l=1}^{L} \cos\left[2\pi(f_0 + \delta f_l)t + \pi k t^2\right] \mathrm{rect}_l(t), \tag{9}$$

$$\mathrm{rect}_n(t) = \mathrm{rect}\left[\left(t - nT_\mathrm{d} - T_\mathrm{d}/2\right)/T_\mathrm{d}\right], \tag{10}$$

where $\gamma = -E_\mathrm{c}^2 \xi \beta_1 \beta_2 V_1 V_\mathrm{RF}/8$. When $\delta = 0$, the center frequency of the IRAJ waveform is identical to that of the adversary radar signal, while its bandwidth is slightly broadened due to the phase modulation, as shown by the blue spectrum in Fig. 1(e-i). When $\delta = 1$, the center frequency of the IRAJ waveform with compound modulation is shown by the pink spectrum in Fig. 1(e-ii). The pulse relationship of the IRAJ waveforms under different modulation modes is shown in Fig. 1(f), where one phase-modulated IRAJ waveform and four compound modulated IRAJ waveforms are arranged in a period. Importantly, the number and timing of these two types of pulses can be flexibly adjusted based on actual requirements.

Then, the IRAJ waveform is amplified by EA2 (Centellax OA4MVM4) and transmitted via a transmitting antenna (TA). Another optical signal from OC3 is sent to PD2 (u2t XPDV2120RA), also to generate the IRAJ waveform, serving as the reference signal for the subsequent de-chirped processing at the radar receiver of the proposed system. The output of PD2 is then amplified by EA3 (Centellax OA4MVM2). At the radar receiver, the echo of the IRAJ waveform reflected by the target is received by the receiving antenna (RA), and the time delay between the echo and the transmitted signal is $\tau$. After being amplified by EA4 (Centellax OA4MVM4), the echo is input to the RF port of a mixer (MIX, COM-MW DM8-18-40G-2537). The reference signal from EA3 is input to the local oscillator (LO) port of this MIX. Thus, the de-chirped signal generated at the intermediate frequency (IF) port of the MIX can be expressed as

$$\begin{aligned} i_\text{de-chirped}(t) \propto\ & \gamma^2 \rho_\mathrm{c}(t,\tau) \sum_{l-1}^{L} \sum_{u=1}^{L} W_{l,u}\left(t,\tau\right) \\ & \times \cos\left\{2\pi\left[k\tau + \delta\left(f_l - f_u\right)\right]t + \varphi_{l,u}\left(\tau\right)\right\} \end{aligned}, \tag{11}$$

$$\rho_c(t,\tau) = \sum_{n=0}^{N-1} \sum_{p=0}^{N-1} c_n c_p \mathrm{rect}_n(t) \mathrm{rect}_p(t-\tau), \tag{12}$$

where $\varphi_{l,u}\left(\tau\right) = 2\pi(f_0 + \delta f_u)\tau - \pi k \tau^2$ stands for the time-delay-dependent phase term of de-chirped signal, and $W_{l,u}\left(t,\tau\right) = \mathrm{rect}_l(t)\mathrm{rect}_u(t-\tau)$ represents the temporal overlap window between the $l$-th frequency-shifting segment of the reference signal and the $u$-th frequency-shifting segment of the echo. Meanwhile, $\rho_{\mathrm{c}}(t,\tau)$ denotes the time-domain product of the binary code sequence in the reference signal and that of the echo. It can be seen that the de-chirped frequency is $f_{\text{de-chirped}} = k\tau + \delta\left(f_l - f_u\right)$. When $\delta = 0$, the de-chirped frequency is determined by the time delay $\tau$. However, due to the binary code sequence $c_n \in \{-1,1\}$, when $\tau$ is close to 0, $\rho_{\mathrm{c}}(t,\tau) \approx 1$ within the temporal overlap between the reference signal and the echo, while it becomes 0 outside the overlap region. In this case, the de-chirped signal is almost unaffected by $\rho_{\mathrm{c}}(t,\tau)$, thus enabling radar detection. Nevertheless, when the time delay exceeds one code duration (i.e., $\tau > T_{\mathrm{d}}$), $\rho_{\mathrm{c}}(t,\tau) = \pm 1$ within the overlap region, and π-phase jumps will be introduced to the de-chirped signal, thus may failing to achieve radar detection. In addition, when $\delta = 1$, the de-chirped signal not only has π-phase jumps, but its de-chirped frequency also becomes $f_{\text{de-chirped}} = k\tau + f_l - f_u$ due to the influence of segmented frequency shifting.

To achieve effective target detection for the proposed IRAJ system, an oscilloscope (OSC, Rohde & Schwarz, RTO2032) is used to acquire the de-chirped signal, and the de-chirped signal corresponding to the IRAJ waveform with only phase modulation (i.e., $\delta = 0$) is selected for processing. This de-chirped signal can be expressed as

$$\begin{aligned} i_{\text{de-chirped}}(t) \propto \gamma^2 \rho_{\mathrm{c}}(t,\tau) \sum_{l-1}^{L} \sum_{u=1}^{L} W_{l,u}\left(t,\tau\right) \\ \times \cos\left\{2\pi k\tau t + 2\pi f_0 \tau - \pi k \tau^2\right\} \end{aligned}. \tag{13}$$

When $0 < \tau < T_{\mathrm{d}}$, the overlap between the reference signal and the echo signal exhibits a high degree of overlap, resulting in few π-phase jumps in the de-chirped signal, thus retaining certain detection capability. However, since $T_{\mathrm{d}}$ is typically short, when the target distance increases, $\tau$ will be much larger than $T_{\mathrm{d}}$. In this way, the $c_n$ of the reference signal and that of the echo signal will be completely misaligned, and a large number of random π-phase jumps will appear. These random jumps are jointly determined by $\tau$ and $c_n$, and the spectrum of the de-chirped signal no longer possesses a single-tone characteristic, but instead exhibits the spectral feature of a pseudo-random binary phase-coded signal.

To eliminate the impact of π-phase jumps on our radar detection, a time-domain squaring operation is performed on the selected de-chirped signal. Ignoring the DC term, the squared de-chirped signal can be expressed as

$$\begin{aligned} i^2_{\text{de-chirped}}(t) \propto \gamma^4 \rho^2{}_{\mathrm{c}}(t,\tau) \sum_{l-1}^{L} \sum_{u=1}^{L} W_{l,u}\left(t,\tau\right) \\ \times \cos\left\{4\pi k\tau t + 4\pi f_0 \tau - 2\pi k \tau^2\right\} \end{aligned}, \tag{14}$$

$$\rho^2{}_{c}(t,\tau) = \sum_{n=0}^{N-1} \sum_{p=0}^{N-1} \mathrm{rect}_n(t)\mathrm{rect}_p(t-\tau). \tag{15}$$

Since the squaring operation results in $\rho^2{}_{\mathrm{c}}(t,\tau) = 1$ within the temporal overlap between the reference signal and the echo, the π-phase jumps in the de-chirped signal can be eliminated. Consequently, the spectrum of the de-chirped signal restores its single-tone characteristic with the de-chirped frequency doubled to $f_{\text{de-chirped}} = 2k\tau$ . Therefore, the target distance can be calculated as follows

$$R = \frac{c\tau}{2} = \frac{cf_{\text{de-chirped}}}{4k}. \tag{16}$$

Through squaring processing, there is no need to pre-knowledge the coding sequence from the transmitter, nor to perform complex demodulation operations at the receiver, which can effectively reduce the signal processing pressure at the receiving end. Limited by the laboratory space, $\tau$ may be less than $T_d$, and the de-chirped signal only has a small number of π-phase jumps. Considering the scenario of long-distance targets, to increase the total time delay between the echo signal and the reference signal, a 200 m SMF is added before PD1 to equivalently simulate the situation where the target range is relatively far.

At the adversary's radar receiver, if a conventional de-chirping reception is adopted, since its reference signal is a pure LFM signal, the de-chirped signal of the adversary's radar can be expressed as

$$\begin{aligned} i_{\text{de-chirped,adv}}(t) \propto \gamma V_1 \text{rect}\left(\frac{t - T_p/2}{T_p}\right) \sum_{n=0}^{N-1} c_n \text{rect}_n(t-\tau) \\ \times \sum_{l=1}^{L} \cos\left[2\pi\left(k\tau - \delta f_l\right)t + \varphi_{l,\text{adv}}\left(\tau\right)\right] \text{rect}_l(t-\tau) \end{aligned}, \tag{17}$$

where $\varphi_{l,\text{adv}}\left(\tau\right) = 2\pi\left(f_0 + \delta f_l\right)\tau - \pi k \tau^2$ is the phase term of the adversary's de-chirped signal. It can be seen that the de-chirped signal will be affected by the combined influence of random binary phase-coded modulation and segmented frequency shifting. A large number of π-phase jumps will be introduced to the adversary's de-chirped signal, thereby degrading its detection performance. Compared with Ref. [20] that employed periodic binary phase modulation, the jamming result of the proposed system do not exhibit regularly spaced intervals in range. It is because, if the binary sequence $c_n$ is a periodic phase sequence, according to the Fourier series expansion, the corresponding frequency-domain expression of the term $\sum_{n=0}^{N-1} c_n \text{rect}_n(t-\tau)$ in (17) becomes a frequency comb. Thus, the time-domain multiplication in (17) corresponds to a frequency-domain convolution, leading to evenly spaced virtual de-chirped peaks. In contrast, the binary sequence $c_n$ used in our system is pseudo-random instead, which effectively randomizes the spectral shifts of the de-chirped signal and introduces noise-like peaks. Besides, when $\delta = 1$ , the adversary's de-chirped signal with noise-like peaks is further shifted by $f_l$, further preventing the adversary's radar from completing target detection.

Similarly, if the adversary's radar adopts pulse compression to process the echo signal, the output of the adversary's matched filter will fail to form a sharp main lobe. This is because binary phase-coded modulation and segmented frequency shifting destroy the correlation between the echo signal and the matched filter, making it impossible for the adversary's matched filter to achieve effective pulse compression.

## 3. Experimental results

A proof-of-concept experiment based on the setup shown in Fig. 1 is performed to verify the feasibility of the proposed system. In the experiment, a fixed pulse relation scheme is adopted. Let m denote the pulse index, where $m = 0,1,2,3,\cdots$. LFM pulses with even pulse indices $m$ are selected to generate phase-modulated IRAJ waveforms, while LFM pulses with odd pulse indices $m$ are selected to generate IRAJ waveforms with superimposed segmented frequency shifting. Limited by the laboratory equipment conditions, AWG1 and AWG2 cannot achieve signal synchronization, making it difficult for the orthogonal RF signals $S_{\text{I-RF}}(t)$ and $S_{\text{Q-RF}}(t)$, generated by AWG2 to be strictly aligned in the time domain with the LFM pulse output from AWG1. Therefore, it is difficult to apply the frequency-shifting modulation to the odd-index LFM pulses as designed. To address this issue, the generation method of the IRAJ waveform is equivalently adjusted. The MZM in the lower branch in Fig. 1 is removed, and two channels of AWG1 are used to separately generate $S'_{\text{I-RF}}(t)$ and $S'_{\text{Q-RF}}(t)$, with the expression of $S'_{\text{I-RF}}(t)$ rewritten as

$$\begin{aligned} S'_{\text{I-RF}}(t) = V_{\text{RF}} \sum_{n=0}^{N-1} c_n \text{rect}_n(t) \\ \times \sum_{l=1}^{L} \cos\left[2\pi\left(f_{\text{RF}} + \delta f_l\right)t\right] \text{rect}_l(t) \end{aligned}. \quad (18)$$

where $f_{\text{RF}}$ stands for its center frequency. Meanwhile, the initial frequency $f_0$ of the LFM signal generated by AWG1 is changed to $f_0' = f_0 - f_{\text{RF}}$. When the pulse index $m$ is even, let $\delta = 0$, and the RF signals are only subjected to pseudo-random binary modulation in the digital domain. When the pulse index $m$ is odd, let $\delta = 1$. Upon phase modulation, the center frequency of the RF signals will be segmentally shifted to $f_{\text{RF}} + f_l$. Subsequently, these two phase-modulated and frequency-shift orthogonal RF signals are input into the RF port of DP-MZM2. The rest of the system shown in Fig. 1 remains unchanged, and finally the designed IRAJ waveform can still be generated.

### 3.1 Generation of IRAJ waveforms

A 16-dBm CW light centered at 1552.525 nm and generated from the LD is used as the light source and injected into DP-MZM1. Then, an LFM signal with a peak-to-peak amplitude of 350 mV is generated by the AWG1 at a sampling rate of 64 Gsa/s. Limited by the maximum signal duration of 4 μs of AWG1 at this sampling rate, the pulse width Td and repetition period $T$ of the LFM signal are set to 1 μs and 2 μs, respectively, and two consecutive LFM pulses form a 4-μs modulation unit. The starting frequency $f_0$ of the LFM signal is set to 4 GHz with its bandwidth $B$ set to 1 GHz, 2 GHz, and 4 GHz, respectively. Subsequently, the 4-μs modulation unit of two LFM pulses is input into DP-MZM1. Another two channels of AWG1 generate two orthogonal RF signals, with a center frequency $f_{\text{RF}}$ of 6 GHz, a signal duration of 1 μs, and a repetition period of 2 μs. When the code length $N$ of the pseudo-random binary phase coding sequence $c_n$ is set to 128, with the number of frequency-shifting segments per pulse $L$ configured to 2, the two segmental frequency shifts are $f_1$=64 MHz and $f_2$=128 MHz, respectively, which satisfy $f_1$=$N$/2 MHz and $f_2$=$N$ MHz numerically. Based on the principle and experimental setup discussed in Section II, the frequency ranges of the IRAJ waveforms with even pulse indices are 10–11 GHz, 10–12 GHz, and 10–14 GHz in sequence, while pulses with odd pulse indices achieve two equal-interval segmented frequency shifts of 64 MHz and 128 MHz. To verify the generation of the IRAJ waveform, the output of PD1 is connected to a spectrum analyzer (Rohde & Schwarz FSP-40) to analyze its electrical spectrum, which is shown in Fig. 2(a).

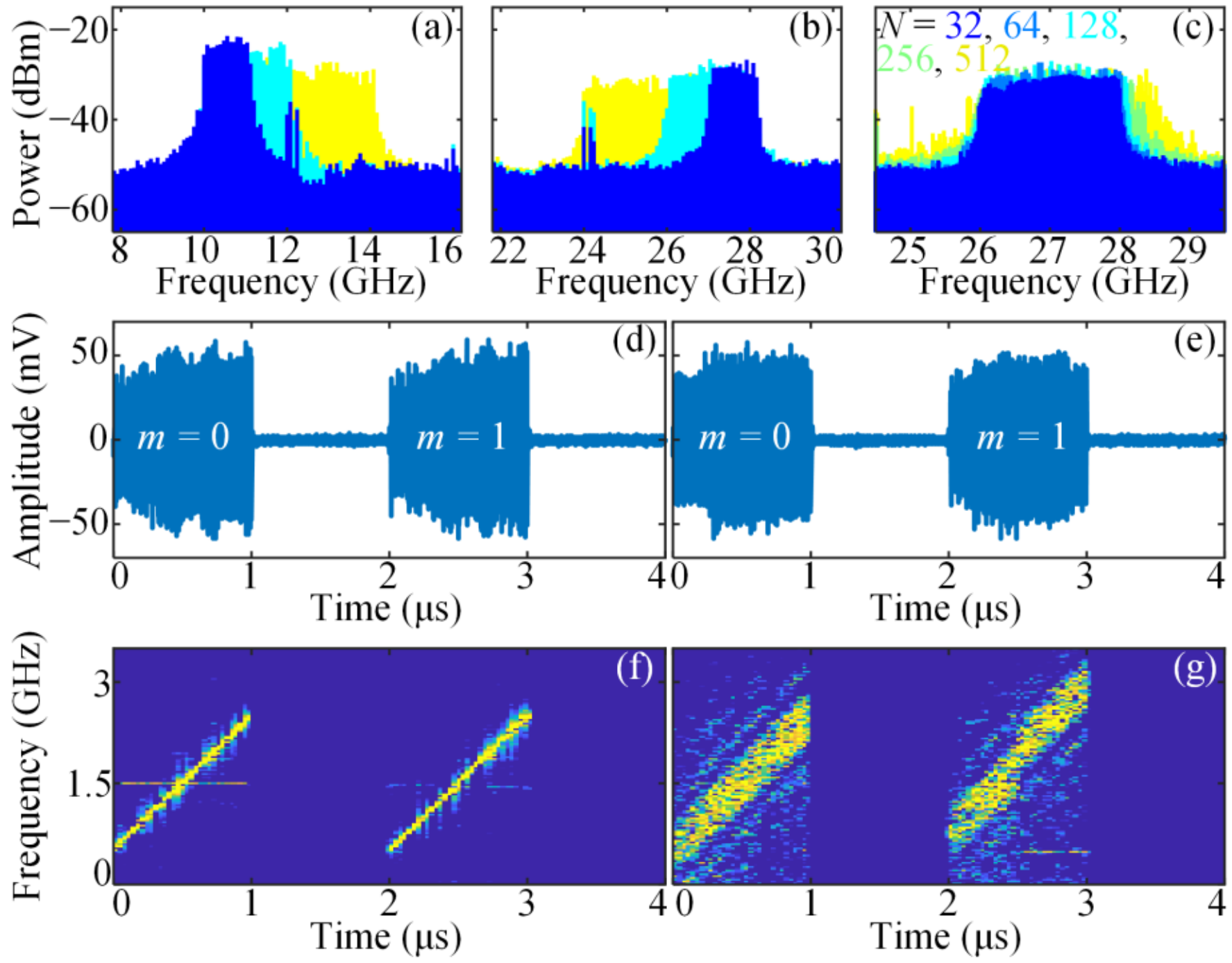

Fig. 2. (a) Spectra of low-frequency radar detection and jamming waveforms with bandwidths of 1, 2, and 4 GHz; (b) spectra of high-frequency IRAJ waveforms with the same bandwidths; (c) spectra of 26–28 GHz IRAJ waveforms with even pulse indices under $N$=32, 64, 128, 256, 512; (d) time-domain waveforms for $N$=32; (e) time-domain waveforms for $N$=512; (f) time-frequency diagrams for IRAJ waveforms with $N$=32 and $L$=2; (g) time-frequency diagrams for IRAJ waveforms with $N$=512 and $L$=2.

In addition, the system can also generate high-frequency IRAJ waveforms with frequency ranges of 27–28 GHz, 26–28 GHz, and 24–28 GHz, which are shown in Fig. 2(b). In this case, the center frequency $f_{\mathrm{RF}}$ of the RF signals is 12 GHz, and the starting frequency $f_0$ of the LFM signal is 15 GHz, 14 GHz, and 12 GHz, respectively. As can be seen from Fig. 2(a) and Fig. 2(b), compared with the power of the low-frequency IRAJ waveforms, the power of the high-frequency IRAJ waveforms is slightly reduced. This is because the AWG1, electrical cables, and other electrical components exhibit greater loss in the high-frequency band. Subsequently, with $L$=2 and the 26–28 GHz IRAJ waveform, $N$ is set to 32, 64, 128, 256, and 512 sequentially. The spectra of these IRAJ waveforms under even pulse indices are shown in Fig. 2(c). As the code length $N$ of the binary phase coding sequence $c_n$ increases gradually from 32 to 512, the spectrum of the IRAJ waveform is significantly broadened. This is because a longer code length $N$ leads to more π-phase jumps in the IRAJ waveform, causing the energy originally concentrated within the LFM bandwidth to spread to the wider frequency ranges.

Fig. 2(d)–(g) further exhibit the time-domain waveforms and time-frequency diagrams of the IRAJ waveforms with pulse indices $m$=0 and $m$=1 under the above conditions. It should be noted that, limited by the maximum sampling rate of 10 Gsa/s and the maximum processing bandwidth of 3 GHz of the adopted OSC, the 26–28 GHz IRAJ waveforms are down-converted to 0.5–2.5 GHz using a 25.5 GHz single-tone signal. Fig. 2(d) and (f) correspond to the time-domain waveforms and time-frequency diagrams when $N$=32, $f_1$=16 MHz and $f_2$=32 MHz, while Fig. 2(e) and (g) correspond to those when $N$=512, $f_1$=256 MHz and $f_2$=512 MHz. As can be seen from the time-frequency diagrams for $m$=0 in Fig. 2(f) and (g), the linear chirp profile gradually becomes less distinct as $N$ increases, and the noise-like components around the profile increase significantly, consistent with the spectrum broadening observed in Fig. 2(c). For $m$=1, when $N$=32, the time-frequency results of the even-index and odd-index pulses in Fig. 2(f) are basically identical due to the small frequency shifts. When $N$=512, because of the increased frequency shifts, there is a significant frequency difference between the time-frequency profiles of the odd-index and even-index pulses in Fig. 2(g). An upward step-like frequency jump occurs at the time-domain boundary (2.5 μs) between the two intra-pulse frequency-shifting segments, demonstrating the realization of equal-interval segmented frequency shifting.

### 3.2 Radar performance of the proposed IRAJ system

To fully verify the detection performance of the designed IRAJ waveform, experiments are carried out around three detection scenarios: Static target range measurement, dynamic target range measurement, and radial velocity measurement. In the radar static target range measurement experiment, a corner reflector is used as the target and placed 3 m away from the antenna pair. For the 24–28 GHz IRAJ waveforms with $L$=2 and $N$=32, 128, 512, respectively, the corresponding de-chirped signals are sampled by the OSC at 1 GSa/s, and the even-indexed pulses are selected and processed for subsequent target detection.

First, following the conventional radar de-chirped processing steps, the selected de-chirped signals are processed with a fast Fourier transform (FFT), and the horizontal coordinates of the spectra are converted to the range axis, yielding the de-chirped results shown in Fig. 3(a)–(c). Due to π-phase jumps in the de-chirped signals, the de-chirped results show the spectral feature of a pseudo-random binary phase-coded signal. When $N$=32, the main lobe of the spectrum in Fig. 3(a) is narrow and centered at 3 m, allowing a rough estimation that the target is located near 3 m. However, the exact distance cannot be determined via the peak detection method. As the code length $N$ increases, the main lobe continues to broaden. When $N$=512, the spectrum of the de-chirped signal in Fig. 3(c) is almost close to the noise floor, making it difficult to identify the approximate range of the target.

To eliminate π-phase jumps in the de-chirped signals and restore the detection capability of the proposed IRAJ system, a time-domain squaring operation is performed on the de-chirped signals, as described in Section II. The processed de-chirped results are shown in Fig. 3(d)–(f). As can be seen, under $N$=32, 128, and 512, the range of the corner reflector is measured to be 3.036 m, 3.035 m, and 3.035 m, respectively, with a range measurement error of approximately 3 cm. This verifies that the scheme based on the squaring operation of de-chirped signals exhibits good effectiveness and stability under different code lengths.

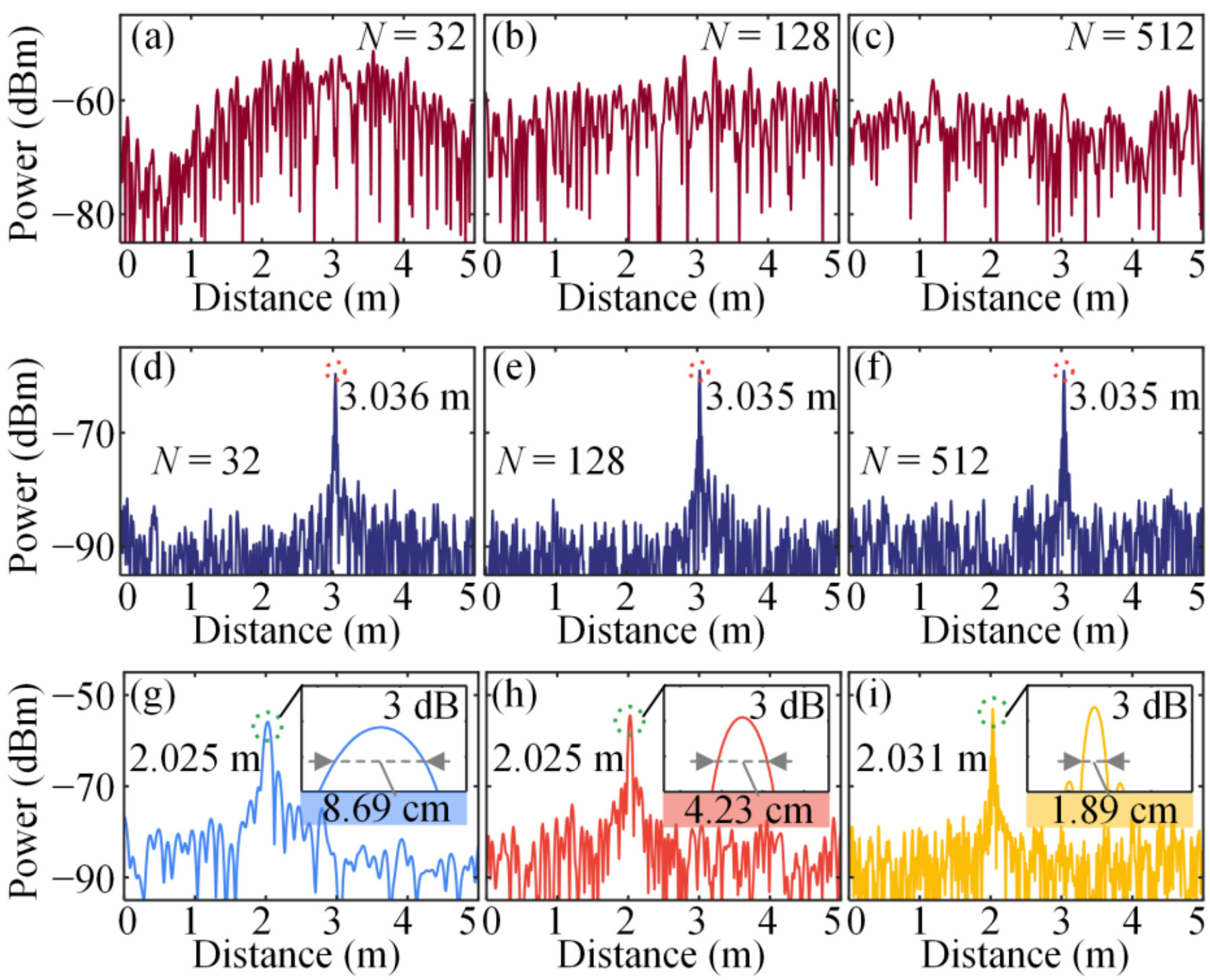

Fig. 3. Static target range measurement results. (a)–(c) Conventional de-chirped results at different code lengths *N*; (d)–(f) Corresponding de-chirped results after squaring operation; (g)–(i) De-chirped results after squaring at different bandwidths of 1, 2, and 4 GHz, respectively.

To further verify the tunability of the system, the bandwidths of the IRAJ waveforms are set to 1 GHz, 2 GHz, and 4 GHz, with the corresponding frequency range being 27–28 GHz, 26–28 GHz, and 24–28 GHz, respectively. Meanwhile, the corner reflector is placed at 2 m. By performing the square operation on the de-chirped signals, the de-chirped results are shown in Fig. 3(g)–(i). It can be seen that the ranging results for these bandwidths are 2.025 m, 2.025 m, and 2.031 m, respectively, with errors generally around 3 cm. In addition, the full width at half maximums (FWHMs) of these de-chirped peaks are 8.69 cm, 4.23 cm, and 1.89 cm, respectively, which are smaller than the theoretical range resolutions of the LFM signals under the corresponding bandwidths: 15 cm, 7.5 cm, and 3.75 cm. The reason for the “narrowed” FWHM of the de-chirped peaks is that the doubled de-chirped frequency after squaring changes the mapping relationship between the range axis and the de-chirped frequency from the original $R = cf_{\text{de-chirped}}/(2k)$ to $R = cf_{\text{de-chirped}}/(4k)$. Therefore, under the same frequency range $\Delta f$, the corresponding range width $\Delta R$ on the original range axis is compressed to $\Delta R/2$. The measured FWHM of the de-chirped peaks after squaring under the three bandwidths should be 17.38 cm, 8.46 cm, and 3.78 cm, respectively.

Subsequently, the IRAJ waveforms are used to dynamic target range measurement. Fig. 4(a) shows the schematic diagram of the dynamic range measurement experiment, where a cylinder is fixed on a turntable. The turntable rotates clockwise with a rotation period of 24.56 s, and its distance from the antenna pair is approximately 1.45 m. In this experiment, the OSC is controlled to acquire the de-chirped signals with specified time intervals to realize the dynamic measurement of the distance between the cylinder and the antenna. When the 24–28 GHz IRAJ waveforms with $N$=128 and $L$=2 are applied, 36 equally spaced sampling points at 1 s intervals over 1.5 rotation periods are measured 10 times. The even-indexed de-chirped signals are selected for the squaring operation. The radar dynamic range measurement results are shown in Fig. 4(b), where the blue circles represent the average measurement values at each sampling point, and the black curve represents the theoretical range variation curve. It can be seen that the measured curve closely matches the theoretical curve. The red crosses in Fig. 4(b) represent the range errors, with the 36th point showing the maximum error of approximately 5.45 cm.

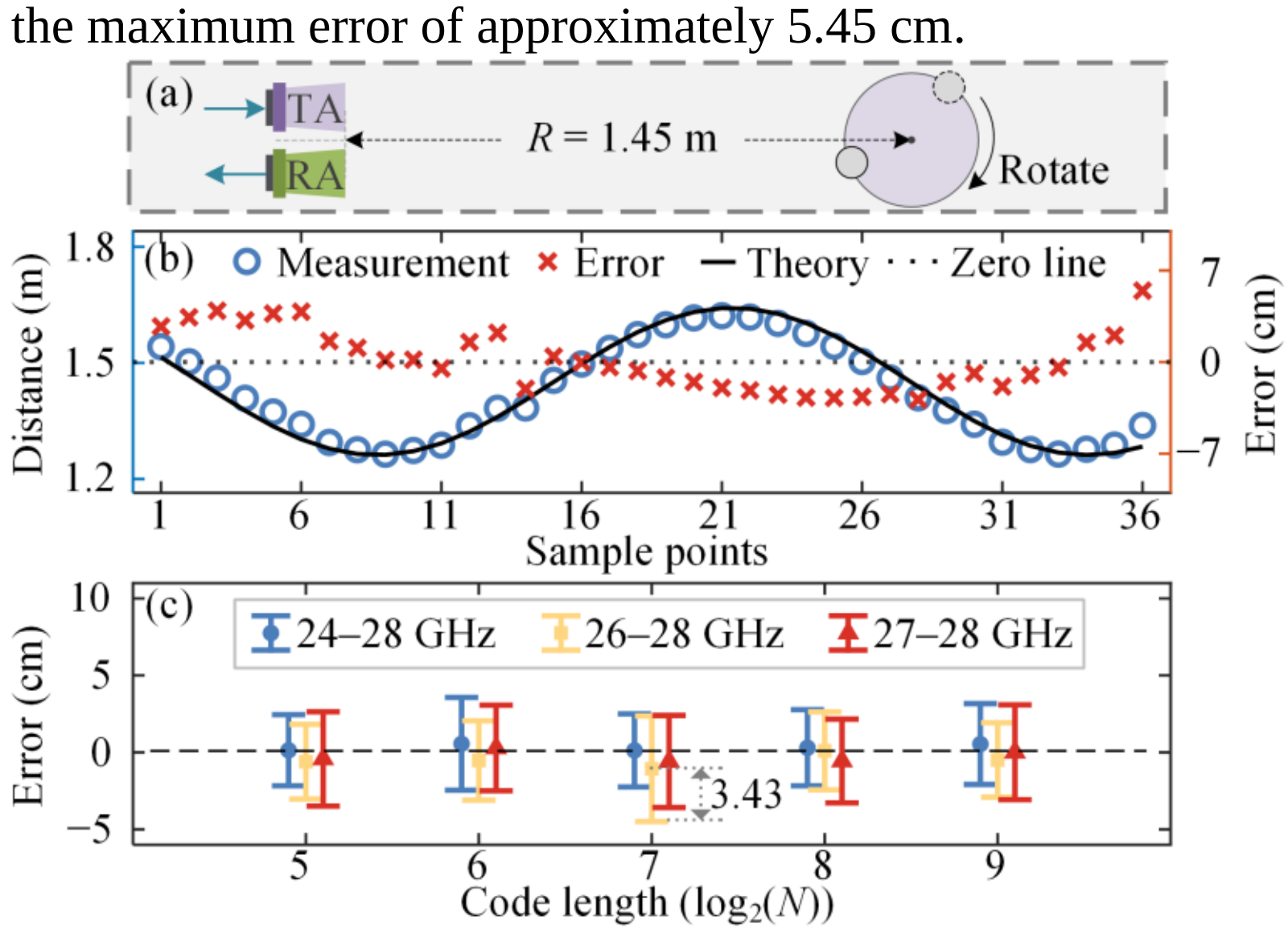

Fig. 4. Dynamic target ranging experiment results. (a) Schematic diagram of the experimental setup; (b) Measured and theoretical ranging results of the 24–28 GHz IRAJ waveform at $N$=128; (c) Statistical ranging errors under different frequency bands and code lengths $N$.

Similarly, the 26–28 GHz and 27–28 GHz IRAJ waveforms are also applied to dynamic target range measurement. The code length $N$ of these IRAJ waveforms with three different bandwidths is varied to 32, 64, 128, 256, and 512, respectively. Error statistics are performed on the 10 repeated measurements of 36 sampling points for each code length, as shown in Fig. 4(c). For clarity of display, the horizontal axis in Fig. 4(c) is labeled in the logarithmic form of the code length $N$. It can be found that the range measurement performance of these IRAJ waveforms under different code lengths is basically stable, with the range measurement error not exceeding ±5cm. In addition, the results of the 26–28 GHz IRAJ waveform exhibit the maximum standard deviation of range measurement error at $N$=128, which is approximately 3.43 cm, which is still within a small error range, indicating that the system has good range measurement stability under different frequency bands and code lengths.

Next, the generated IRAJ waveforms are employed for radial velocity measurement. Fig. 5(a) shows the schematic diagram of the radial velocity measurement scheme. Due to the lack of a suitable target for uniform linear motion, a turntable-based alternative scheme is adopted for radial velocity measurement. A cylinder is placed on the turntable, with the center of the turntable slightly offset from the center of the antenna pair. When the turntable rotates clockwise or counterclockwise at an angular velocity $\omega$, the velocity direction of the cylinder will be aligned with the radar line-of-sight direction for a certain time. By measuring the target's distances $L_1$ and $L_2$ at two extremely close time instants with an interval of $\Delta t$, the radial velocity of the target can be calculated based on the rate of change in range with respect to time.

In the experiment, the rotation period of the turntable is set to 24.56 s with its angular velocity $\omega$ of 0.256 rad/s. The rotational radius of the cylinder is 60.5 cm. Therefore, the theoretical radial velocities of the cylinder when rotating clockwise and counterclockwise are 15.48 cm/s and −15.68 cm/s, respectively. The time interval $\Delta t$ is set to 0.3 s, and the 24–28 GHz and 26–28 GHz IRAJ waveforms are applied to radar radial velocity measurement, with the code length $N$ set to 32, 64, 128, 256, and 512 in sequence. The statistical results of 5 measurements during clockwise and counterclockwise rotation are shown in Fig. 5(b) and (c), respectively. It can be seen that the mean values of the measured radial velocities under both bandwidths fluctuate around the theoretical values. The maximum standard deviations of the radial velocities in the clockwise and counterclockwise directions are under the 2 GHz bandwidth IRAJ waveforms, with 3.46 cm/s and 3.77 cm/s, respectively.

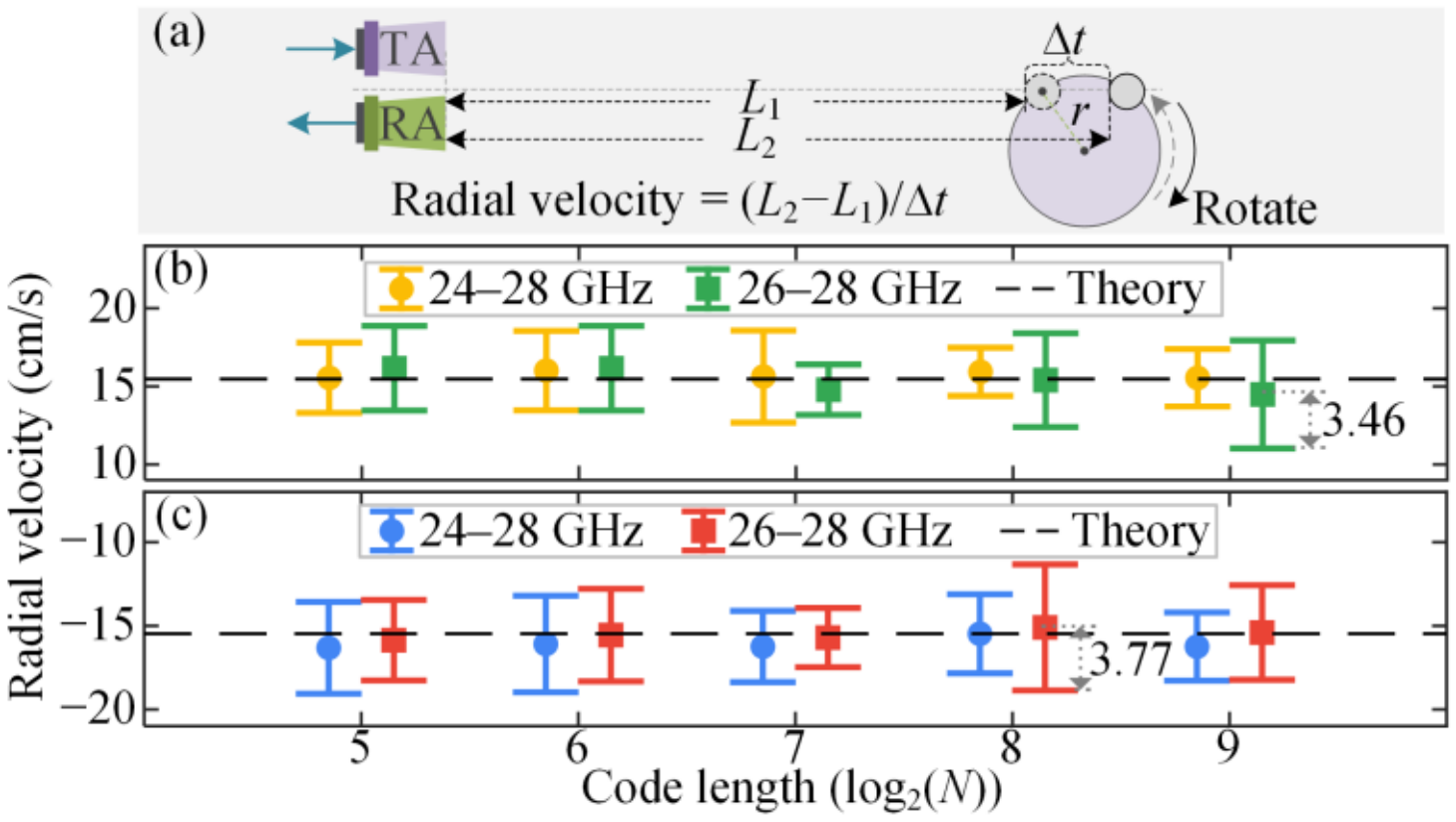

Fig. 5. Radial velocity measurement results. (a) Schematic diagram of the radial velocity measurement scheme; (b) and (c) Statistical velocity results under different frequency bands and code lengths $N$ during clockwise and counterclockwise rotation, respectively.

### 3.2 Jamming performance against adversary radar

To verify the jamming performance of the proposed IRAJ waveforms against the adversary radar adopting de-chirped reception, the signal at the LO port of the MIX at the radar receiver is replaced with an LFM signal generated by AWG1, which serves as the adversary reference signal. The LFM signal has a sweep range of 24–28 GHz, a pulse width $T_p$ of 1 μs, and a pulse repetition period $T$ of 2 μs. After the IRAJ waveform is input into the RF port of the MIX, the de-chirped signal from the IF port of the mixer is captured by the OSC and then processed using FFT in the digital domain. By converting the frequency axis to the range axis, the de-chirped results are shown in Fig. 6(a)–(d), where Fig. 6(a) represents the de-chirped result of the adversary radar without jamming. It can be seen from Fig. 6(a) that when the target is placed at 2 m, the spectrum of the de-chirped signal without jamming has an obvious single-tone peak, indicating that the adversary can achieve accurate radar ranging at this time, with a measured distance of 2.016 m.

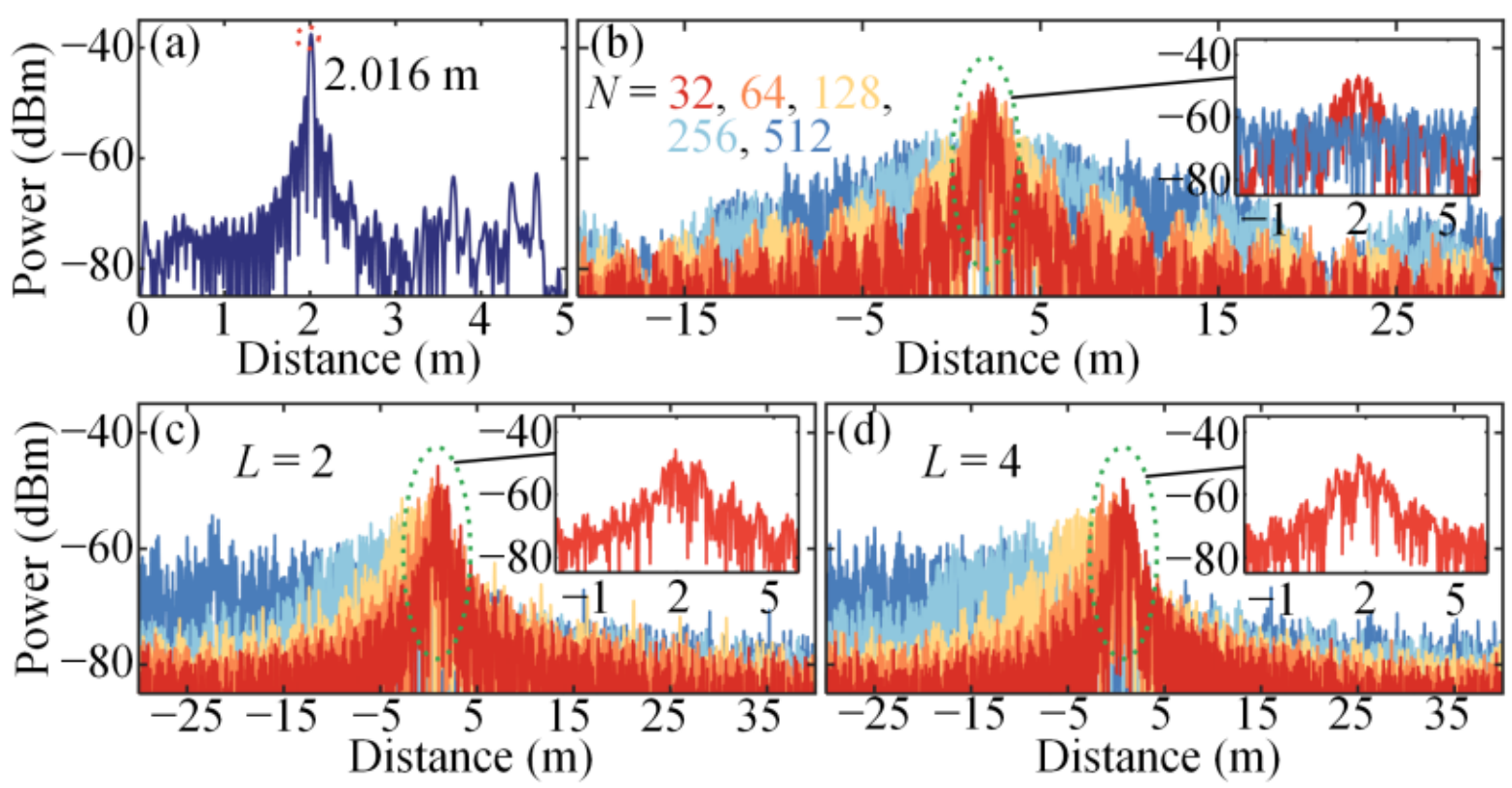

Fig. 6. De-chirped results of the adversary radar. (a) De-chirped result without jamming. (b) De-chirped results under jamming by 24–28 GHz even-indexed IRAJ waveforms where $N$=32, 64, 128, 256, and 512 from red to dark blue. (c) and (d) De-chirped results under jamming by 24–28 GHz odd-indexed IRAJ waveforms with $L$=2 and $L$=4, respectively.

The de-chirped results obtained at the adversary receiver when receiving the IRAJ waveform are shown in Fig. 6(b)–(d). First, de-chirped results corresponding to the even-indexed pulses are analyzed, as shown in Fig. 6(b), where red, orange, yellow, light blue, and dark blue correspond to the cases when code length $N$ of the IRAJ waveform is set to 32, 64, 128, 256, and 512, respectively. It can be seen that the peak at the target distance completely disappears, and the spectrum exhibits the spectral shape of the pseudo-random binary code sequence, centered at the target distance of 2 m. The insets in Fig. 6(b) are the locally enlarged spectra near the target distance when $N$=32 and $N$=512. As the code length $N$ increases, the main lobe width of the spectrum gradually broadens, making the spectrum near the target distance closer to the noise floor. Therefore, the adversary cannot achieve radar ranging through the conventional peak detection method, and the target range information is effectively concealed.

The de-chirped results corresponding to the odd-indexed pulses are further analyzed. Fig. 6(c) and (d) show de-chirped results when the number of frequency-shifting segments $L$ is set to 2 and 4. The frequency shifts are designed based on the code length $N$ and the number of frequency-shift segments $L$. For $L$=2, the frequency shifts are $f_1$=$N$/2 MHz and $f_2$=$N$ MHz. For $L$=4, the frequency shifts are $f_1$=$N$/2 MHz, $f_2$=$N$ MHz, $f_3$=3$N$/2 MHz, and $f_4$=2$N$ MHz. Specifically, when $N$=128, the frequency shifts for $L$=2 are $f_1$=64 MHz and $f_2$=128 MHz, while for $L$=4, they are $f_1$=64 MHz, $f_2$=128 MHz, $f_3$=192 MHz, and $f_4$=256 MHz. Compared with the symmetric spectrum corresponding to the even-indexed pulses in Fig. 6(b), the de-chirped spectrum of the odd-indexed pulses is shifted to the left and no longer centered around the target distance. This phenomenon becomes more pronounced as $N$ increases, since a longer code length results in a larger frequency shift, causing a more noticeable shift in the spectrum. Even if the adversary attempts to roughly estimate the target position through the spectral symmetry, it cannot be achieved, and the concealment effect of the distance information is more thorough. In addition, with the same code length $N$, as $L$ increases from 2 to 4, the spectrum of the de-chirped signal is significantly broadened, thus a broadband noise-like jamming is formed against the adversary.

In addition to the de-chirped reception, pulse compression is another commonly used method for processing radar signals. To further evaluate the jamming applicability of the proposed IRAJ waveform, the following experiments focus on its impact on the adversary radar employing pulse compression for target ranging. Different from the de-chirped reception method, the time-domain waveform of the radar echo is required to be completely captured to implement pulse compression processing with a corresponding matched filter in the digital domain. Since the maximum sampling rate of the used OSC is 10 GSa/s and the processing bandwidth is only 3 GHz, it is difficult to directly acquire the radar echo. Therefore, down-conversion of the echo is required. A 25.5 GHz single-tone signal generated by a microwave signal generator (Agilent 83630B) serves as an LO signal for down-converting the 26–28 GHz IRAJ waveform to 0.5–2.5 GHz, making it easier for acquisition by the OSC. In the digital domain, an LFM signal with a sweep range of 0.5–2.5 GHz is constructed as the matched filter. After implementing frequency-domain matched filtering, the range measurement results at the receiver of the adversary radar based on pulse compression are obtained in Fig. 7(a)–(d).

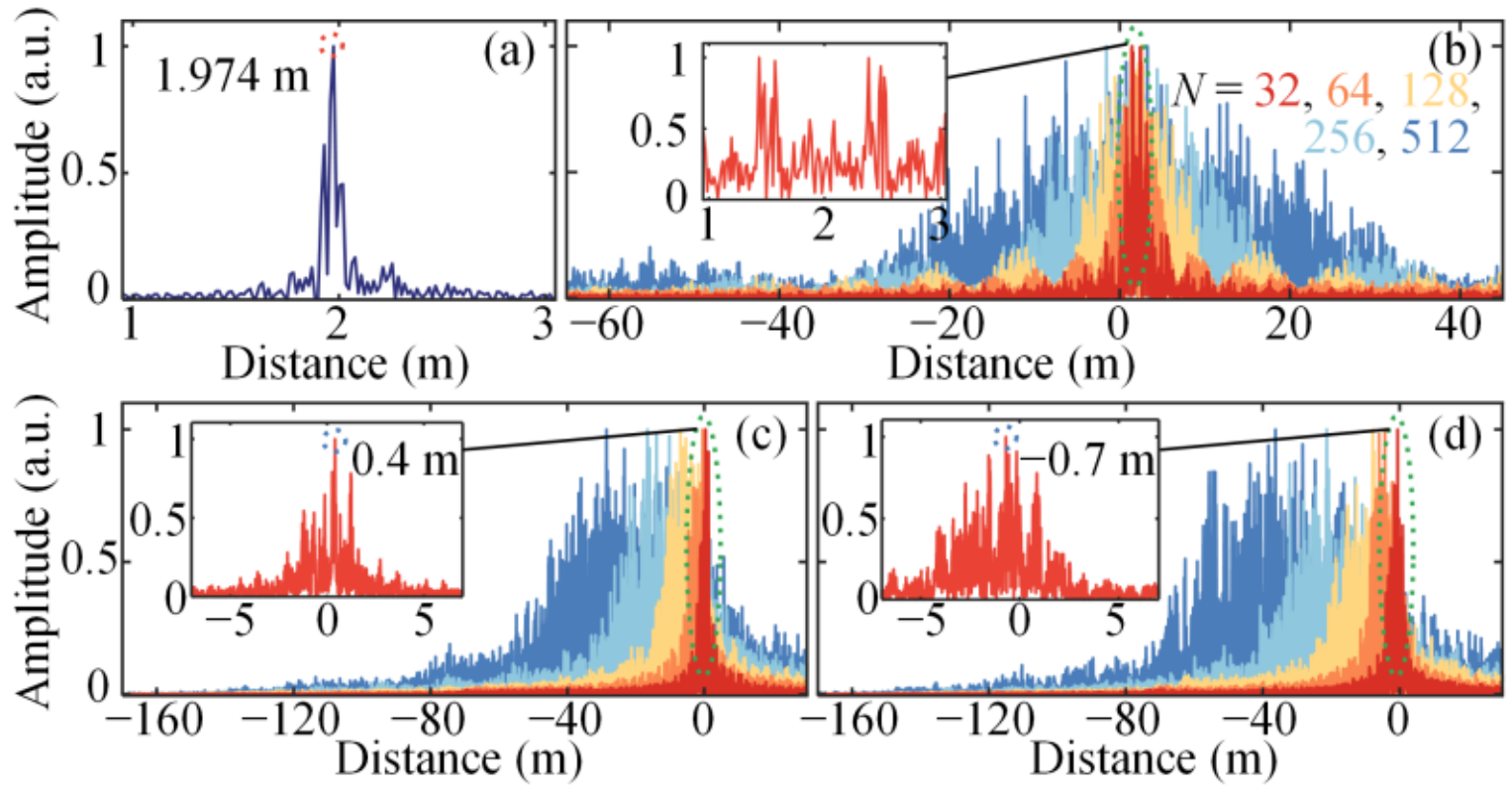

Fig. 7. Pulse compression results of the adversary radar. (a) Pulse compression result without jamming. (b) Pulse compression results under jamming by 26–28 GHz even-indexed IRAJ waveforms where $N$=32, 64, 128, 256, and 512 from red to dark blue. (c) and (d) Pulse compression results under jamming by 26–28 GHz odd-indexed IRAJ waveforms with $L$=2 and $L$=4, respectively.

Fig. 7(a) shows the range measurement result of the 2 m target with no jamming. It can be seen that an obvious pulse compression peak appears at the target distance, with a measured distance of 1.974 m, indicating that the adversary radar can achieve accurate ranging. Fig. 7(b) shows the pulse compression results at the adversary radar receiver after receiving the IRAJ waveforms with merely binary phase-coded modulation, where the code length $N$ is set to 32, 64, 128, 256, and 512. Similar to the results when the adversary adopts de-chirped reception, the peak originally at the target distance disappears. This is because the binary phase-coded modulation destroys the linear phase relationship of the LFM signal and the matched filter, causing the autocorrelation gain to spread rather than concentrate into a single peak. Moreover, as $N$ increases, the main lobe width gradually broadens. However, the pulse compression results are still centered at the target distance. The adversary can roughly infer the target interval through the symmetry feature, which poses a certain risk of jamming failure.

The jamming effect of the IRAJ waveforms with superimposed segmented frequency shifting is then verified. Fig. 7(c) and (d) show the pulse compression results for $L$=2 and $L$=4, respectively. Due to the additional frequency offsets introduced by the segmented frequency shifting, the pulse compression result is no longer centered at the target distance, but is shifted to the left. As $L$ increases, for the same code length $N$, the main lobe width of the pulse compression result further expands, also resulting in a noise-like jamming. As shown in the enlarged insets in Fig. 7(c) and (d), when the code length $N$ is 32, and $L$ is 2 and 4, respectively, the peaks of the pulse compression results are at 0.4 m and −0.7 m, causing an increase in ranging error. The adversary is unable to perform conventional ranging by locating the pulse compression peak or even inferring the approximate target position from the symmetry of the pulse compression result.

## 4. Discussion

### 4.1 Tunability of the IRAJ system

To further verify the adaptability of the proposed IRAJ system to adversary radar signals across different frequency bands, the frequency range of the LFM signal generated by AWG1 is adjusted, thereby altering the frequency of the generated IRAJ waveforms. As shown in Fig. 2(a), the system can generate IRAJ waveforms in the lower frequency bands of 10–11 GHz, 10–12 GHz, and 10–14 GHz. First, a dynamic target range measurement experiment is carried out with the IRAJ waveform in the low-frequency band. A cylinder with a rotation period of 30.11 s is fixed on the turntable, whose distance from the antenna pair is adjusted to 1 m, and the oscilloscope is controlled to acquire de-chirped signals at 1 s intervals. With the code length $N$ set to 32, 64, 128, 256, and 512, and the frequency-shifting segments $L$ set to 2, de-chirped signals of the 10–14 GHz IRAJ waveform are acquired at 36 equally spaced sampling points within 1.2 rotation periods. The even-indexed de-chirped signals are then selected for squaring processing. The dynamic range measurement results when $N$=128 are shown in Fig. 8(a), where the measured distance curve composed of blue circles is highly consistent with the theoretical curve. The red crosses represent the range measurement errors at each sampling point, with a maximum error of approximately 1.64 cm, smaller than the maximum error of 5.45 cm in the 24–28 GHz band. This is because the distance of the cylinder in the low-frequency band is closer to the antenna pair, resulting in a higher echo signal power, which improves the signal-to-noise ratio (SNR) at the receiver of our radar and thus enhances the ranging performance.

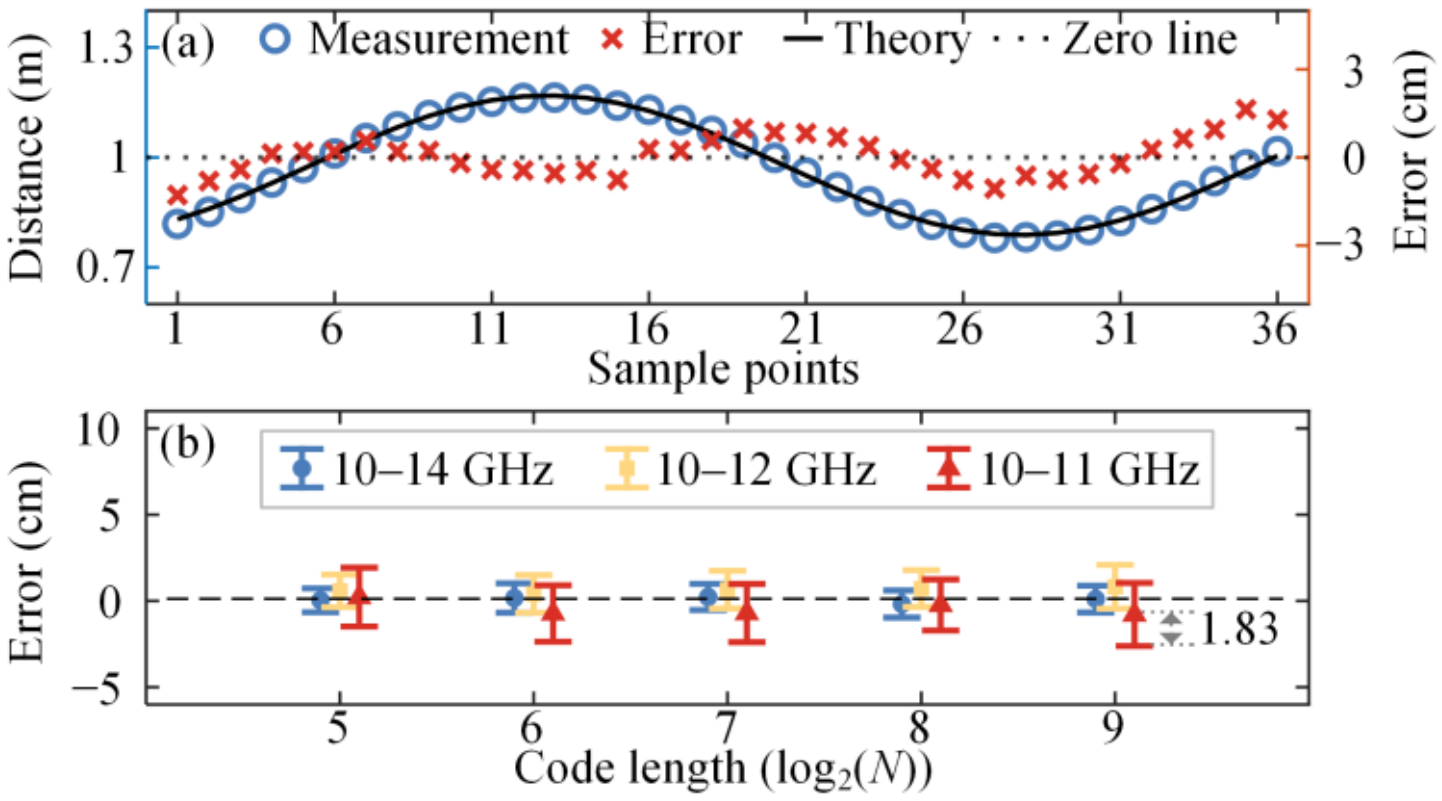

Fig. 8. Dynamic ranging results for low-frequency IRAJ waveforms. (a) Measured and theoretical ranging results of the 10–14 GHz IRAJ waveform at $N$=128; (b) Statistical ranging errors under different frequency bands and code lengths.

Then the de-chirped signals of the other two IRAJ waveforms with frequency ranges of 10–11 GHz, and 10–12 GHz, are also collected 36 times The error statistics results of these IRAJ waveforms with different frequencies are shown in Fig. 8(b), where the blue dots, yellow square and red triangle represent the mean values of range measurement errors for each frequency range, and the length of the error bars reflects the standard deviation of the measurement results. It can be seen that the mean values of the ranging errors are distributed near 0. The maximum standard deviation of ranging errors is approximately 1.83 cm at code length $N$=512 of the 10–11 GHz IRAJ waveform. This indicates that the IRAJ waveforms in low-frequency bands with different bandwidths can all achieve accurate range measurement.

Then, the 10–14 GHz and 10–12 GHz IRAJ waveforms are applied to radial velocity measurement. The experimental setup and parameters are the same as those used for the high-frequency situation, with the theoretical velocity being 15.48 cm/s and −15.68 cm/s for clockwise and counterclockwise rotation, respectively. The corresponding results are shown in Fig. 9. The mean measured velocities for both bandwidths fluctuate around the theoretical values. In addition, the velocity measurement results of the 10–14 GHz IRAJ waveform exhibit a smaller standard deviation compared with those of the 10–12 GHz IRAJ waveform, whose maximum velocity measurement error standard deviation is 3.60 cm/s for clockwise rotation and 3.66 cm/s for counterclockwise rotation, respectively. The results confirm that the proposed IRAJ system achieves reliable across-band tunability for target detection, with measurement errors staying within acceptable limits in both high and low frequency ranges.

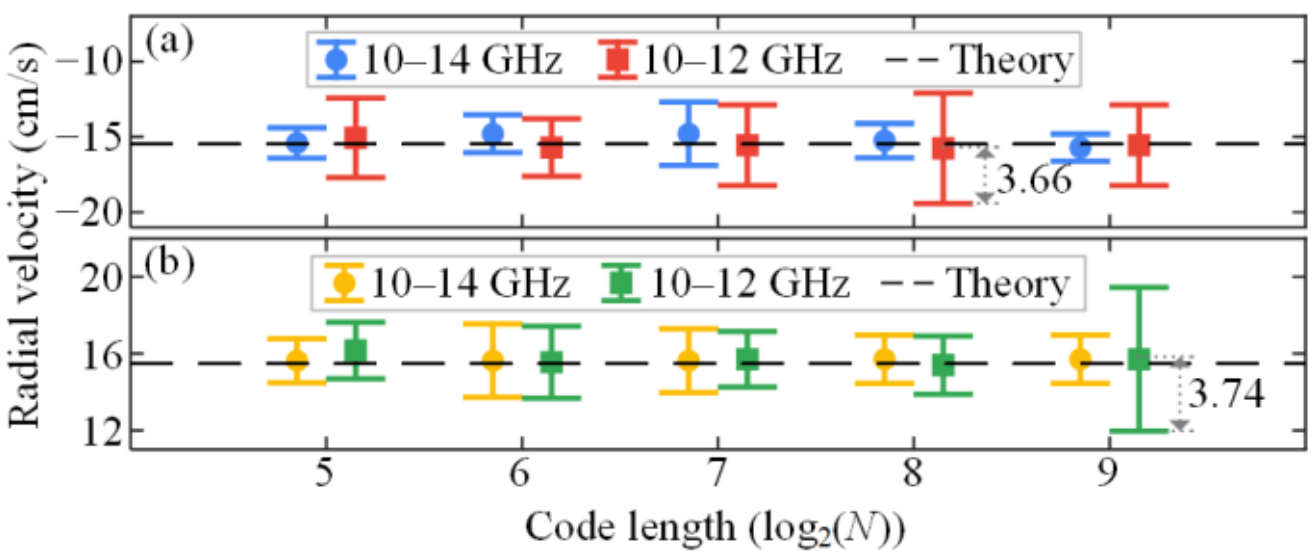

Fig. 9. Radial velocity measurement results for low-frequency IRAJ waveforms. (a) and (b) Statistical velocity results under different frequency bands and code lengths during counterclockwise and clockwise rotation, respectively.

After completing the functional verification of the target detection capability of the low-frequency IRAJ waveform, the following analyzes its jamming effect against adversary radars. The corner reflector is placed at 2 m. Fig. 10(a) and (c) show the ranging results of the adversary radar without jamming, where the echo of the 10–14 GHz LFM signal is processed by the de-chirped reception, and the radar echo of the 10–12 GHz LFM signal is processed by pulse compression, respectively. It can be seen that the measured distances under these two situations are 2.029 m and 2.038 m, respectively, indicating that the adversary can achieve radar ranging using both radar signal processing methods without jamming. Fig. 10(b) and (d) show the range measurement results at the adversary receiver when receiving the designed IRAJ waveforms in the corresponding frequency bands. The code length $N$ of the IRAJ waveforms is set to 32, 64, 128, 256, and 512, and the number of frequency-shifting segments $L$ is set to 4. Similar to the results in the high-frequency situation, regardless of whether the adversary uses de-chirp reception or pulse compression, the peaks in the ranging results of adversary radar disappear, showing a disordered noise-like profile. Moreover, the main lobe width broadens significantly as the code length $N$ increases. When an appropriate code length $N$ and frequency shifts are selected, the IRAJ waveform can form broadband noise-like jamming against the adversary radar, causing the range measurement of adversary radar to fail to work properly.

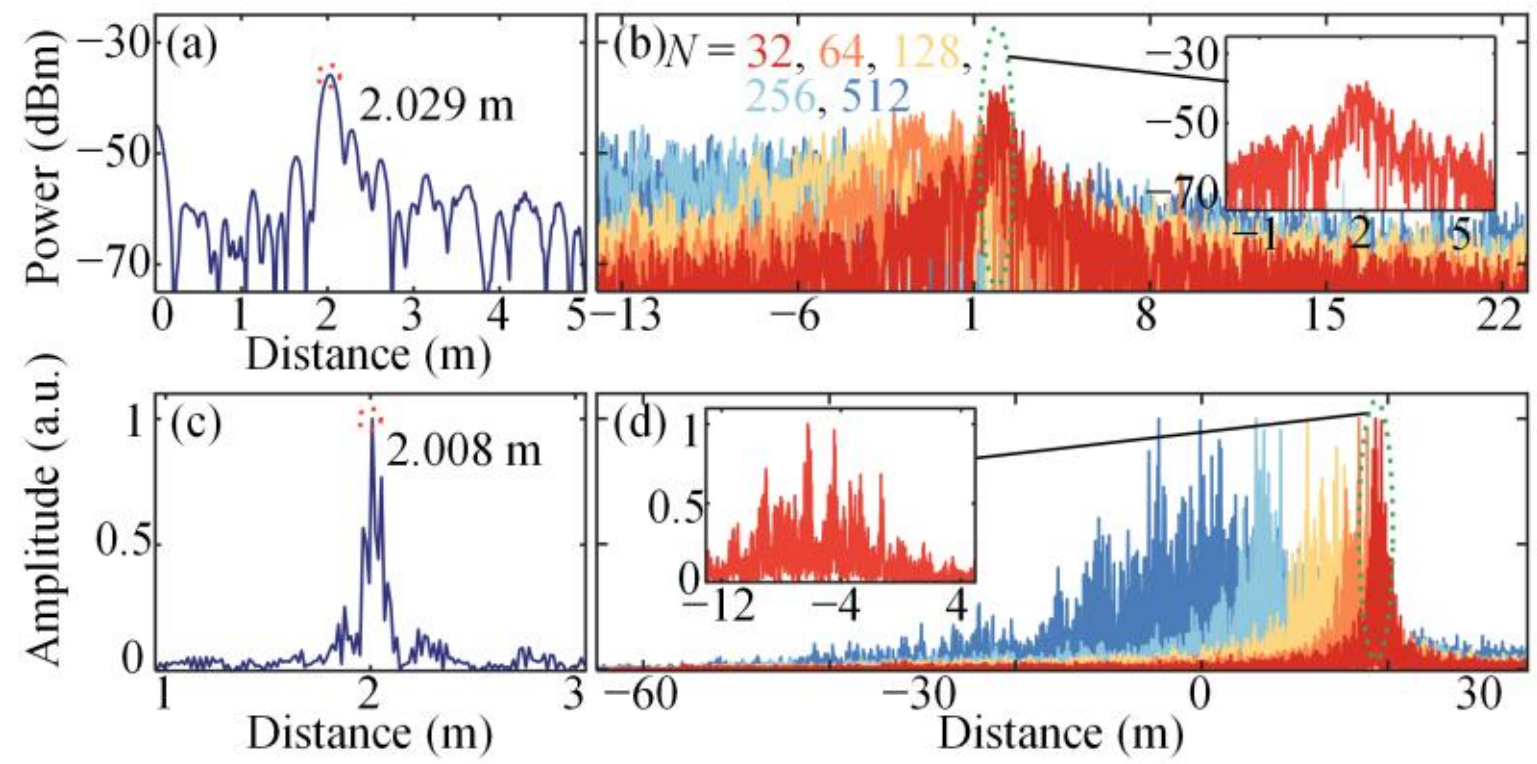

Fig. 10. Jamming performance of low-frequency IRAJ waveforms. (a) De-chirped ranging result of 10–14 GHz LFM signal without jamming; (b) De-chirped results under jamming by 10–14 GHz odd-indexed IRAJ waveforms with $L$=4, where code lengths $N$=32, 64, 128, 256, and 512 from red to dark blue. (c) Pulse compression ranging result of 10–12 GHz LFM signal without jamming; (d) Pulse compression spectra under jamming by 10–12 GHz odd-indexed IRAJ waveforms with $L$=4.

It should be noted that the frequency range for the generated IRAJ waveforms can cover a broader tuning range, with the limitations imposed only by the working frequency of certain devices, including the antenna pair of 10–40 GHz, and PDs of 0–30 GHz. For demonstration purposes, two frequency ranges of 10–14 GHz and 24–28 GHz are selected as examples in the experiments, serving as practical evidence of the system's flexible tunability, which can be adjusted based on the specific requirements of the warfare scenarios.

### 4.2 Comparison with existing methods

A comparison between the proposed microwave photonic IRAJ system and previously reported electrical IRAJ systems, together with photonics-assisted single-function systems, is given in Table I.

As can be seen, the advantages and key contributions of this work are summarized as follows: (1) To the best of our knowledge, this is the first microwave photonic system that simultaneously implements precise radar detection and coherent suppressive jamming. Most existing microwave photonic schemes focus on either radar or jamming separately, while our system enables the two functions to operate concurrently using a IRAJ waveform. (2) Compared with most previously reported electronic IRAJ systems, which are limited by fixed frequency bands and narrower bandwidths, the proposed IRAJ system supports a bandwidth of up to 4 GHz with flexible frequency coverage across 10–28 GHz. (3) Different from most existing microwave photonic jamming schemes that generate symmetric and equally spaced false targets, the proposed system realizes noise-like jamming against both de-chirped reception and pulse compression radars with noise-like ranging results, improving jamming effectiveness from being identified and countered.

Despite the above advantages, the proposed microwave photonic IRAJ system still has some limitations. In practice, the adversary radar signal inevitably enters our receiver, thus introducing additional interference when performing the squaring operation for radar detection capability. To avoid degrading the target detection performance under strong adversary signal power conditions, a spatial-domain isolation based on phased-array beamforming can be employed to suppress the incoming adversary radar signal before it affects the receiver.

**Table 1. Comparison with previously reported works**

| | IRAJ system | Center frequency /Bandwidth | Radar jamming | | Experimental validation |
|---|---|---|---|---|---|
| | | | Jamming format | Jamming distribution | |
| Ref. [20] | No | 10.5 GHz/ 300 MHz, 1 GHz | PSJ[1] | Symmetric, equally spaced | Yes |
| Ref. [22] | No | 10, 30 GHz/1 GHz | CMSJ, ISRJ | Symmetric, equally spaced | Yes |
| Ref. [24] | No | 7.5 GHz/1 GHz | CSMJ, C&IJ | Asymmetric, equally spaced | Yes |
| Ref. [27] | Yes | 1 GHz/50 MHz | PSJ | noise-like | No |
| Ref. [28] | Yes | 1 GHz/5, 20MHz | FSJ[2], TDJ[3] | Single | No |
| This work | Yes | 12, 26 GHz/ 1, 2, 4 GHz | FSJ, PSJ | noise-like | Yes |

[1] Phase-shifting jamming, [2] Frequency-shifting jamming, [3] Time-delay jamming.

## 5. Conclusion

In this work, we have proposed and experimentally demonstrated a microwave photonic IRAJ system based on pseudo-random binary modulation and segmented frequency shifting. The system can concurrently realize target detection and effective jamming, with a 4 GHz instantaneous bandwidth and flexible frequency coverage across 10–28 GHz. Unlike conventional microwave photonic jamming schemes that generate easily identifiable symmetric and equally spaced false targets, the designed IRAJ waveform produces noise-like jamming results, significantly enhancing anti-identification capability. Besides, using the squaring operation without prior coding sequence knowledge, accurate target sensing ability with a ranging error of around 5 cm, and radial velocity measurement error below 4 cm/s. To our best knowledge, this is the first experimentally verified microwave photonic IRAJ system achieving concurrent high-performance detection and noise-like jamming, which breaks through the electronic bottleneck in bandwidth and frequency tunability, and fills the gap in shared waveform design and joint signal processing for integrated microwave photonic radar and jamming systems.

## Acknowledgements

National Natural Science Foundation of China (62371191); Shanghai Oriental Talent Program (QNJY2024007); Science and Technology Commission of Shanghai Municipality (22DZ2229004).